\documentclass[notitlepage,aps,amsmath,amssymb,graphicx,onecolumn]{revtex4-1}

\usepackage{dcolumn}
\usepackage{bm}
\usepackage{multirow}
\usepackage{float} 
\usepackage[dvipdfmx]{graphicx} 
\usepackage{threeparttable}
\usepackage{here}
\usepackage[top=30truemm,bottom=30truemm,left=25truemm,right=25truemm]{geometry}

\begin{document}

\title{Calculation of the Residual Entropy of Ice Ih \\
by Monte Carlo simulation with the Combination of the Replica-Exchange Wang-Landau algorithm and
Multicanonical Replica-Exchange Method}

\author{Takuya Hayashi$^{1}$\thanks{tahayashi@tb.phys.nagoya-u.ac.jp}, 
Chizuru Muguruma$^{2}$\thanks{muguruma@lets.chukyo-u.ac.jp}, and
Yuko Okamoto$^{1,3,4}$\thanks{okamoto@tb.phys.nagoya-u.ac.jp}}
\affiliation{
$^{1}$Department of Physics, Graduate School of Science, Nagoya University, Nagoya, Aichi 464-8602, Japan \\
$^{2}$Faculty of Liberal Arts and Sciences, Chukyo University, Nagoya, Aichi 466-8666, Japan \\
$^{3}$Center for Computational Science, Graduate School of Engineering, Nagoya University, Nagoya, Aichi 464-8603, Japan \\
$^{4}$Information Technology Center, Nagoya University, Nagoya, Aichi 464-8601, Japan
}




\begin{abstract}
We estimated the residual entropy of ice Ih by the recently developed simulation protocol, namely, 
the combination of Replica-Exchange Wang-Landau algorithm and Multicanonical Replica-Exchange Method.
We employed a model with the nearest neighbor interactions on the three-dimensional hexagonal lattice,
which satisfied the ice rules in the ground state.
The results showed that our estimate of the residual entropy is found to be within $0.038$ \% of series expansion estimate by Nagle 
and within $0.000077$ \% of PEPS algorithm by Vanderstraeten.
In this article, we not only give our latest estimate of the residual entropy of ice Ih
but also discuss the importance of the uniformity of a random number generator in MC simulations.
\end{abstract}

\pacs{Valid PACS appear here}

\maketitle
\section{Introduction}
After the experimental discovery that the ice Ih has non-zero residual entropy near zero temperature \cite{ICE_GIAUQUE},
the theoretical explanation about the origin was proposed by the ice rules \cite{ICE_BERNAL}, 
which considered the hydrogen bonds between water molecules in ice \cite{ICE_PAULING}. 
The residual entropy per one water molecule $S_{0}$ is proportional to the logarithm of the number of 
degrees of freedom of the
 orientations of one water molecule $W_{0}$: 
\begin{equation}
 \label{EQ1}
S_{0} = k_{\rm B} \ln{W_{0}}.
\end {equation}
Here, $k_{B}$ is the Boltzmann constant and the value is $1.9872 {\rm \left[ cal/deg \right]}$.
The estimate by Pauling was $W_{0}^{{\rm Pauling}}=1.5$ and $S_{0}^{{\rm Pauling}} \simeq 0.806$ \cite{ICE_PAULING}.
It was in accord with the experimental value $S_{0}^{{\rm Experiment}} = 0.82 (5)$ \cite{ICE_GIAUQUE}.
Error bars in this article are given with respect to the last digits in parentheses.
However, it was shown that the Pauling's estimate was a lower bound by Onsager and Dupuis \cite{ICE_ONSAGER} 
and the advanced theoretical approximation was obtained by Nagle \cite{ICE_NAGLE}.

As for computational simulations, two simulation models (2-state model and 6-state model), 
which satisfied the ice rules in the ground state, were proposed and the value was estimated \cite{ICE_BERG_2007,ICE_BERG_2007_2,ICE_BERG_2008,ICE_BERG_2012}
by the {\it Multicanonical} (MUCA) {\it Monte Carlo} (MC) {\it Method} \cite{MUCA1,MUCA2} (for reviews, see, e.g., \cite{MUCA3,MUCA_BOOK}). 
After these simulation models were suggested, many research groups estimated 
the residual entropy by various computational approaches for the last decade (see, e.g., 
\cite{ICE_HERRERO,ICE_KOLAFA,ICE_FERREYRA1,ICE_FERREYRA2,ICE_VANDERSTRAETEN}).
The estimates by computer simulations seem to be equal to or more accurate than theoretical estimate by Nagle.
Although the residual entropy of ice is becoming one of good examples to test the accuracy of simulation algorithms,
there seem to be small disagreements among the estimates.
The exact residual entropy of Ice Ih has yet to be obtained.

In this article, we present our latest estimate of the residual entropy by the recently proposed MC simulation 
with the combination of {\it Replica-Exchange Wang-Landau algorithm} (REWL) \cite{REWL1,REWL2}
and {\it Multicanonical Replica-Exchange Method} (MUCAREM) \cite{MUCAREM1,MUCAREM2,MUCAREM3},
which we refer to as REWL-MUCAREM \cite{REWL-MUCAREM}.
REWL-MUCAREM can give us high precise estimates of the density of state (DOS) and the entropy under appropriate computational conditions.
We employed the 2-state model proposed in \cite{ICE_BERG_2007}.
Our latest result is in good agreement with the estimates by several research groups which used other simulation methods.
In addition, we also report that the uniformity of the random numbers is important for MC simulations.

This article is organized as follows.
We summarize the results of previous researches briefly in Sec. II.  
In Sec.~III, we explain the ice model that we employed and the REWL-MUCAREM protocol.
In Sec.~IV, the simulation details are given. 
In Sec.~V, our results are presented, and Sec.~VI is devoted to conclusions.
In Appendix A, the importance of random numbers is discussed.

\section{Residual entropy}
Figure~\ref{STR_2D} shows the hexagonal crystal structure of ice Ih in two-dimensional projections. 
Figures~\ref{STR_2D}(a) and \ref{STR_2D}(b) correspond to the projection to the $xy$-plane and the $yz$-plane, respectively.
We assume that the water molecules exist as $\rm H_{2}O$ molecules in ice and
hydrogen atoms can occupy one of the two places on each bond according to the ice rules in \cite{ICE_PAULING}: 
(1) there is one hydrogen atom on each bond, and (2) there are two hydrogen atoms near each oxygen atom.

Suppose that there are $N$ water molecules.
The number of hydrogen atoms is $2N$.
The theoretical residual entropy $S_{0}$ per one water molecule is defined by:
\begin{equation}
 \label{E2}
 S_{0} = \displaystyle \frac{k_{{\rm B}} \ln W}{N} = k_{{\rm B}} \ln W_{0},
\end{equation}
where 
\begin{equation}
 \label{E2_1}
  W=(W_{0})^{N}.
\end{equation}
Here, $W$ is the total number of configurations of water molecules which satisfy the two ice rules.
By defining $W_{0}$ as the number of orientations per one water molecule, Pauling estimated the value to be \cite{ICE_PAULING}
\begin{equation}
 \label{E3}
 \displaystyle
  W_{0}^{{\rm Pauling}} = 1.5.
\end{equation}
His strategy is as follows: ignoring the second ice rule (two hydrogen atoms exist near each oxygen atom), $2^{2N}$ configurations can be considered 
because each hydrogen atom is given the choice of two positions on each bond.
There are $16$ arrangements of the four hydrogen atoms around one oxygen atom and the only $6$ arrangements can satisfy the second ice rule. 
Thus, the total number of configurations $W$ that satisfies the ice rule (1) and ice rule (2) simultaneously is:
\begin{equation}
 \label{E4}
 \displaystyle
  W =( W_{0}^{\rm Pauling})^{N} = 2^{2N} \times \left( \frac{6}{16} \right)^{N} = \left( \frac{3}{2} \right)^{N}.
\end{equation}
Eq. (\ref{E4}) can be converted to the residual entropy as
\begin{equation}
 \label{E5}
 \displaystyle
 \begin{split}
  S_{0}^{{\rm Pauling}} &= k_{\rm B} \ln \left(W_{0}^{{\rm Pauling}} \right) \\
                              &= 0.80574 \cdots \, {\rm \left[ cal / deg~mole \right]}.
 \end{split}
\end{equation}

Onsager and Dupuis showed that $W_{0}^{\rm Pauling}=1.5$ is in fact a lower bound because Pauling's arguments omitted the effects of closed loops~\cite{ICE_ONSAGER}.
Nagle used a series expansion method in order to refine the theoretical estimate~\cite{ICE_NAGLE}.
The contribution coming from short closed loops were taken into account counting the graph of the loops directly and
the effects of long loops were estimated by extrapolation based on the results of short loops.
The approximate value was
\begin{equation}
 \label{E6}
 \displaystyle
  W_{0}^{{\rm Nagle}} = 1.50685(15),
\end{equation}
and
\begin{equation}
 \label{E7}
 \displaystyle
  S_{0}^{{\rm Nagle}} = 0.81480(20) \cdots \, {\rm \left[ cal / deg~mole \right]}.
\end{equation}
Here, the error bar is not statistical but reflects higher-order corrections of the expansion,
 which are not entirely under control. 
In terms of theoretical approximation, another series expansion method, which used 
numerical linked cluster (NCL) expansion, were proposed \cite{ICE_SINGH}.

With the development of computer science, 
many research groups have tried to estimate the residual entropy by various computational approach
(for example, Thermodynamic Integration method, Wang-Landau algorithm, and PEPS algorithm) \cite{ICE_HERRERO,ICE_FERREYRA1,ICE_FERREYRA2,ICE_KOLAFA,ICE_VANDERSTRAETEN}.
However, there remain small differences between these results.
We give our latest estimate by REWL-MUCAREM protocol in this article.

\section{Models and Methods}
 \subsection{Models}
We used the 2-state model \cite{ICE_BERG_2007}. 
In this model, we do not consider distinct orientations of the water molecule (the ice rule~(2) is ignored), 
but allow two positions for each hydrogen nucleus between two oxygen atoms (the ice rule~(1) is always satisfied). 
The total potential energy $E$ of this system is defined by
\begin{equation}
 \label{E8}
 \displaystyle
  E=-\sum_{i} f \left(i,b_{i}^{1},b_{i}^{2},b_{i}^{3},b_{i}^{4} \right),
\end{equation}
where $i$ stands for a site number of oxygen atoms.
The sum is over all sites (oxygen atoms) of the lattice.
The function $f$ is given by
\begin{equation}
 \label{E9}
 \displaystyle
  f \left(i,b_{i}^{1},b_{i}^{2},b_{i}^{3},b_{i}^{4} \right) = \begin{cases}
    2 & {\rm for \, two \, hydrogen \, nuclei \, close \, to \,} i \\
    1 & {\rm for \, one \, or \, three \, hydrogen \, nuclei \, close \, to \,} i \\
    0 & {\rm for \, zero \, or \, four \, hydrogen\, nuclei \, close \, to \,} i
  \end{cases}
\end{equation}
The ground state of this model fulfills the two ice rules completely. 
The energy at the ground state $E_{\rm ground}$ is $-2N$.
Because the normalization (the total number of configurations $\sum_{E} n(E)$ is $2^{2N}$ where $n(E)$ 
is the number of states at energy $E$) is known, 
MUCA simulations allow us to estimate the number of the configurations at the ground state accurately 
by calculating the ratio of $\tilde{n}(E_{\rm ground})$ to $\sum_{E} \tilde{n}(E)$ \cite{PUTTS_BERG_1992}.
Here, $\tilde{n}(E)$ is the estimates obtained from MUCA simulations.

\subsection{Methods}
We used an advanced generalized-ensemble MC algorithm that we recently developed,
REWL-MUCAREM \cite{REWL-MUCAREM}.
In this protocol, the multicanonical weight factor (i.e., the inverse of the DOS) is determined roughly by a REWL simulation
and then the weight factor is refined by repeating MUCAREM simulations.

A brief explanation of MUCA \cite{MUCA1,MUCA2,MUCA3,MUCA_BOOK} is now given here. 
The multicanonical probability distribution of potential energy $P_{{\rm MUCA}}(E)$ is defined by
\begin{equation}
 \label{E10}
 P_{{\rm MUCA}}(E) \propto  g(E) W_{{\rm MUCA}}(E) \equiv {\rm const}~,
\end{equation}
where $W_{{\rm MUCA}}(E)$ is the multicanonical weight factor, 
the function $g(E)$ is the DOS, and $E$ is the total potential energy.
By omitting a constant factor, we have 
\begin{eqnarray}
 \label{E11}
 \displaystyle
 W_{{\rm MUCA}}(E) &= \displaystyle{\frac{1}{g(E)}}~.
\end{eqnarray}
In MUCA MC simulations, the trial moves are accepted with the following Metropolis transition probability
 $w \left( E \rightarrow E' \right)$:
\begin{eqnarray}
 \label{E12}
 w \left( E \rightarrow E' \right) = {\rm min} \sl \left[ 1, \displaystyle \frac{W_{{\rm MUCA}} \sl (E')}
                                                                             {W_{{\rm MUCA}} \sl (E)} \right] 
                                             = {\rm min} \sl \left[ 1, \displaystyle \frac{g(E)}{g(E')} \right]~.
\end{eqnarray}
Here, $E$ is the potential energy of the original configuration and $E'$ is that of a proposed one.
After a long production run, the best estimate of DOS can be obtained 
by the single-histogram reweighting techniques \cite{SHRT}:
\begin{eqnarray}
 \label{E13}
g(E)= \displaystyle \frac{H(E)}{W_{{\rm MUCA} }\sl (E)}~,
\end{eqnarray}
where $H(E)$ is the histogram of sampled potential energy. 
Practically, the $W_{{\rm MUCA}} (E)$ is set to $\exp [-\beta E ]$ at first and 
modified by repeating sampling and reweighting. Here, $\beta$ is the inverse of temperature $T$ 
($\beta = 1/ k_{{\rm B}} T$).

The {\it Wang-Landau} (WL) algorithm \cite{WL1,WL2} also uses $1/g(E)$ as the weight factor 
and the Metropolis criterion is the same as in Eq.~(\ref{E12}). 
However, $g(E)$ is updated dynamically as $g(E) \rightarrow f \times g(E)$ 
during the simulation 
when the simulation visits a certain energy value $E$. $f$ is a modification factor. 
We continue the updating until the histogram $H(E)$ becomes flat. 
If $H(E)$ is flat enough, a next simulation begins after resetting 
the histogram to zero and reducing the modification factor (usually, $f \rightarrow \sqrt{f}$). 
The flatness evaluation can be done in various ways. 
This process is terminated when the modification factor attains a predetermined 
value $f_{{\rm final}}$,
 and  $\exp(10^{-8}) \simeq 1.000\,000\,01$ is often used as $ f_{{\rm final}}$.
Hence, the estimated $g(E)$ tends to converge to the true DOS of the system
within this much accuracy set by $f_{{\rm final}}$. 

MUCA can be combined with {\it Replica-Exchange Method} (REM) \cite{REM1,REM2,MHRT1} for more efficient sampling.
(REM is also referred to as {\it Parallel Tempering} \cite{WHAM1}.)
The method is referred to as MUCAREM \cite{MUCAREM1,MUCAREM2,MUCAREM3}.
In MUCAREM, the entire energy range of interest $\left[ E_{{\rm min}},E_{{\rm max}} \right]$ 
is divided into $M$ sub-regions, $ E_{{\rm min}}^{\{m\}} \leq E \leq E_{{\rm max}}^{\{m\}}$ 
$(m=1,2,\cdots,M)$, where  
$E_{{\rm min}}^{\{1\}} = E_{\rm min}$ and $E_{{\rm max}}^{\{M\}}=E_{\rm max}$. 
There should be some overlaps between the adjacent regions.
MUCAREM uses $M$ replicas of the original system. 
The weight factor for sub-region $m$ is defined by \cite{MUCAREM1,MUCAREM2,MUCAREM3}: 
\begin{eqnarray}
 \label{E15}
    W^{\{m\}}_{{\rm MUCA}} (E) = 
  \displaystyle{
   \begin{cases}
     e^{- \beta_{\rm L}^{\{m\}} E},      & {\rm for} \, \, E < E^{\{m\}}_{{\rm min}} \,\,\,\, ,  \\ 
     \displaystyle{\frac{1}{g_m(E)}},   & {\rm for} \, \, E^{\{m\}}_{{\rm min}} \le E \le E^{\{m\}}_{{\rm max}} 
\,\,\,\, ,  \\
     e^{- \beta_{\rm H}^{\{m\}} E},      & {\rm for} \, \, E > E^{\{m\}}_{{\rm max}} \,\,\,\, ,  
   \end{cases}
  }
\end{eqnarray}
where $g_m(E)$ is the DOS for 
$E^{\{m\}}_{{\rm min}} \le E \le E^{\{m\}}_{{\rm max}}$ in sub-region $m$,
$\beta_{\rm L}^{\{m\}} = d k_{\rm B} \ln \left[g_m(E) \right] / d E ~(E=E^{\{m\}}_{\rm min})$
and, $\beta_{\rm H}^{\{m\}} = d \ln \left[ \it g_m(E) \right] / d E ~(E=E^{\{m\}}_{\rm max})$.
The MUCAREM weight factor $W_{{\rm MUCAREM}}(E)$ for the entire energy range is expressed by the following formula:
\begin{eqnarray}
 \label{E16}
 W_{{\rm MUCAREM}}(E) =     & \displaystyle{ \prod_{m=1}^{M} } W_{{\rm MUCA}}^{\{m\}}(E)~.
\end{eqnarray}
After a certain number of independent MC steps, replica exchange is proposed 
between two replicas, $i$ and $j$, in neighboring sub-regions, $m$ and $m+1$, respectively. 
The transition probability, $w_{{\rm MUCAREM}}$, of this replica exchange is given by
\begin{eqnarray}
 \label{E17}
  w_{{\rm MUCAREM}} &=& \displaystyle 
        {\rm min} \sl \left[ 1 , \frac{W^{\{m\}}_{\rm MUCA}(E_j) W^{\{m+1\}}_{\rm MUCA}(E_i)}  
                                      {W^{\{m\}}_{\rm MUCA}(E_i) W^{\{m+1\}}_{\rm MUCA}(E_j)}  \right] ,
\end{eqnarray}
where $E_i$ and $E_j$ are the energy of replicas $i$ and $j$ before the replica exchange,
respectively.
If replica exchange is accepted, the two replicas exchange their weight factors 
$W^{\{m\}}_{{\rm MUCA}} (E)$ and  
$W^{\{m+1\}}_{{\rm MUCA}} (E)$   
and energy histogram $H_{m}(E)$ and $H_{m+1}(E)$. 
The final estimate of DOS can be obtained from $H_{m}(E)$ after a long production simulation 
by the multiple-histogram reweighting techniques \cite{MHRT2,MHRT3} 
or weighted histogram analysis method (WHAM) \cite{MHRT3}.
Let $n_{m}$ be the total number of samples for the $m$-th energy sub-region.
The final estimate of DOS, $g(E)$, is obtained by solving the following WHAM equations 
self-consistently by iteration \cite{MUCAREM1,MUCAREM2,MUCAREM3}:
\begin{eqnarray}
 \label{E18}
   \left\{
   \begin{array}{l}
      g(E) = \frac{ \displaystyle{ \sum_{m=1}^{M} } H_{m} (E) }
              {\displaystyle{ \sum_{m=1}^{M}} n_{m} \exp \left( f_{m} \right) W_{{\rm MUCA}}^{\{m\}} (E)} ~, \\ \\
      \exp \left( - f_{m} \right) =\displaystyle{ \sum_{E} g(E) W_{{\rm MUCA}}^{\{m\}}(E) }~.
   \end{array}
   \right.
\end{eqnarray}
Repeating these MUCAREM sampling and WHAM reweighting processes can obtain more accurate DOS. 
Although ordinary REM is often used to obtain the first estimate of DOS in the MUCAREM iterations, 
we used the results of REWL simulation \cite{REWL1,REWL2} instead of the first REM run because
REWL is stable and it can give more accurate DOS.

The REWL method is essentially based on the same weight factors as in MUCAREM, while the WL simulations 
replace the MUCA simulations for each replica.
This simulation is terminated when the modification factors on all sub-regions attain a 
certain minimum value $f_{{\rm final}}$. 
After a REWL simulation, $M$ pieces of DOS fragments with overlapping energy intervals are obtained. 
The fragments need to be connected in order to determine 
the final DOS in the entire energy range $\left[ E_{{\rm min}} ,E_{ {\rm max}} \right]$. 
The joining point for any two overlapping DOS pieces is chosen 
where the inverse microcanonical temperature 
$\beta \left(= \partial S(E) /\partial E \right)$ coincides best \cite{REWL1,REWL2}.
This connecting process can be omitted in REWL-MUCAREM because the estimated DOS from WHAM 
is used directly as multicanonical weight factor in MUCAREM.
After repeating MUCAREM several times, the DOS with highest accuracy is obtained.
In this article, ordinary MUCA simulations were performed after REWL-MUCAREM for estimating the errors.

\section{Computational details}
The total number of water molecules $N$ is given by $n_{x} \times n_{y} \times n_{z}$, 
where $n_{x},n_{y}$, and $n_{z}$ are the numbers of sites along the $x,y,$ and $z$ axis, respectively (see Fig.~\ref{STR_2D}).
The total number of sites (i.e., total number of oxygen atoms) is $N$ and the total number of hydrogen atoms is $2N$.
The values of $n_{x}$, $n_{y},$ and $n_{z}$ are restricted to $n_{x}=1,2,3, \cdots$, $n_{y}=4,8,12, \cdots$, 
and $n_{z}=2,4,6, \cdots$, because we used periodic boundary conditions (PBC).
The total number of molecules considered was $N=128, 288, 360, 576,896,1600,2880$, and $4704$.
The positions of hydrogen atoms are updated during MC simulations.
Physical values were collected after each MC step.
One MC sweep is defined as an evaluation of Metropolis criterion $2N$ times.

The REWL-MUCAREM protocol was used in order to obtain the DOS.
It corresponds to the number of configuration $n(E)$ at $E$.
In MUCA and WL MC simulations, it is necessary to determine the entire energy range $[E_{\rm min},E_{\rm max}]$ 
before starting simulations.
We selected the values as follows: $[E_{\rm min},E_{\rm max}]=[-2N,- 5N/4]$.
Here, $E_{\rm min}$ corresponds to the ground state and $E_{\rm max}$ corresponds to the energy value
around which the entropy takes the maximum value (see Fig.~\ref{TPC_ENT}). 
Figure~\ref{TPC_ENT} shows the typical dimensionless entropy ($\ln n(E)$) per one water molecule in ice Ih, 
which was estimated by our additional simulation for the system $N=2880$ under the condition $[E_{\rm min},E_{\rm max}]=[-2N,0]$.
The dimensionless entropy takes the maximum value at $E/N=-5/4$.
Thus, the inverse temperature $\beta$ takes the value $0$ at $E_{\rm max}$.
Under the condition $[E_{\rm min},E_{\rm max}]=[-2N,- 5N/4]$, 
Flat MUCA probability distribution is realized in $-2N \le E \le - 5N/4$ and
canonical probability distribution at $\beta=0$ is obtained in $E > -5N/4$.
The $n(E)$ is summed up to the maximum energy which obtained during simulations 
in order to estimate the total number of configurations.
Although it is desirable to take $E_{\rm max}=0$ in order to estimate $W_{0}$ with high accuracy
according to our normalization, 
it is sufficient that $E_{\rm max}$ is $-5N/4$ because most of the configurations are 
distributed around $E=-5N/4$ and the number of configurations 
which take much higher potential energy than $E=-5N/4$ can be ignored (see Fig. \ref{TPC_NOS}).
Figure~\ref{TPC_NOS} shows the summation of $n(E)$ which was normalized at $E_{\rm min}$ 
per one water molecule for the system $N=128$.
It was summed up from $E_{\rm min}$ to $E$ ($E$ is a certain energy value).
The summation is saturated at a bit larger energy than $E/N=-5/4$.
The difference between the asymptotic value and the value $16/6$, which is the inverse value of Pauling's estimate $6/16$, represents the effects of closed loops in \cite{ICE_ONSAGER}.
The inset in Fig.~\ref{TPC_NOS} shows the $n(E)$ directly.
Most of the total number of configurations are distributed around the peak.
These results implies that the sum of the number of states which takes much higher energy than $E/N=-5/4$ is 
small sufficiently not to affect on our estimates of residual entropy.
In fact, although we compared the estimate of $W_{0}$ under the condition $[-2N,-5N/4]$ with the estimate 
under the condition $[-2N,0]$ up to the $N=2880$ system, the difference was small enough within errors.
As a result for $[E_{\rm min},E_{\rm max}]=[-2N,- 5N/4]$, 
we could obtain more samples at ground energy state, which was the
most important if for the estimation of the residual entropy, during MUCA simulations.

In the REWL and MUCAREM simulations, $4$ to $32$ replicas were used depending on the number of water molecules.
Each replica performed a WL simulation in REWL or a MUCA simulation in MUCAREM  
within their energy sub-regions, which had an overlap of about $80$ \% between neighboring sub-regions.
The replica exchange criterion and WL flatness criterion were tested during the simulations. 
The intervals for replica exchange and flatness tests depend on the lattice sizes (see TABLE~I).
In the WL flatness criterion of each replica, a flatness of $H_{\rm min} / H_{\rm max} > 0.5$ was considered 
sufficient for stopping the recursion and restart a next WL iteration 
by the recursion factor $f \rightarrow \sqrt{f}$. 
Here, $H_{\rm min}$ is the smallest and $H_{\rm max}$ is the largest value of the histogram $H(E)$.
We iterated the reducing $f$ process $20$ times and we set $f_{\rm final} \simeq 1.90735 \times 10^{-6}$. 
Once a rough estimate of DOS was obtained by REWL,
MUCAREM samplings and WHAM reweighting processes were then repeated $5$ times in order to get more precise DOS.
The total number of MC sweeps for each MUCAREM was $2.0 \times 10^{7}$ sweeps.

After we obtained a DOS by REWL-MUCAREM, MUCA production runs were performed $M=32$ times independently 
for evaluating the residual entropy and errors.
Average values and errors were obtained by the following standard formulae:
\begin{eqnarray}
 \label{E_ERROR}
\overline{n(E_{\rm min})} = \frac{ \displaystyle{\sum_{i=1}^{M}} n(E_{\rm min})^{\left\{ i \right\}} }{M} ~~,~~~~~
  \varepsilon_{n} = \sqrt{\frac{
\displaystyle \sum_{i=1}^{M}  \left( n(E_{\rm min})^{\left\{ i \right\}} - 
\overline{n(E_{\rm min})} \right)^{2} }{M \left( M-1 \right)}} \rule[0mm]{0mm}{8mm}.
\end{eqnarray}
Here, $n(E_{\rm min})^{\left\{ i \right\}}$ is a measured value from the $i$-th simulation 
$(i=1,2,\cdots,M)$.
The total number of MC sweeps for measurement was $6.4 \times 10^{8}$ sweeps for each MUCA production run.
The single-histogram reweighting techniques were employed in order to obtain estimates for $W_{0}$. 

Random number generators have a large effect on the MC method (see Appendix A).
In this article, the Mersenne Twister random number generator was employed \cite{MERSENNE1}.
We used the program code on open source \cite{MERSENNE_CODE}.

\section{Results and Discussion}
Figure~\ref{MCRM_ERANK} shows the time series of the energy-range index of one of the replicas (Replica 1) 
during the final MUCAREM simulation for the $N=4704$ system.
Here, we used $32$ replicas.
The total energy range $[E_{\rm min},E_{\rm max}]$ was divided into 32 sub-regions.
$E_{\rm min}$ was $-9408$ and $E_{\rm max}$ was $-5880$.
The minimum energy label was $1$ and the maximum energy label was $32$.
It can be seen that replica $1$ went from label $1$ to label $32$ and came back many times.
This means that replica exchange worked properly.
Figure~\ref{MCRM_E} shows the time series of potential energy of one of the replicas (Replica 1) for the same simulation as in Fig.~\ref{MCRM_ERANK}.
The replica made a random walk in energy space.
There is a strong correlation between energy label in Fig.~\ref{MCRM_ERANK} and the potential energy in Fig.~\ref{MCRM_E}, as expected.
The four figures in Fig.~\ref{MCRM_HIST} show the histograms
 of potential energy which were obtained 
by the final MUCAREM simulation for the $N=4704$ system.
Each energy label corresponds to the sub-region $m$.
Although we used $32$ sub-regions, the only four sub-regions ($m=1,2,3,4$) are shown in Fig.~\ref{MCRM_HIST}.
Each histogram shows a flat distribution.

Figure~\ref{MUCA_ENT} shows the logarithm of our final DOS by the REWL-MUCAREM protocol
for $N=4704$ and Fig. \ref{MUCA_HIST} shows the energy histogram obtained after the MUCA production runs 
which used the final DOS as the weight factor.
The ideal MUCA weight factor makes a completely flat histogram.
The flatness ($H_{\rm min} / H_{\rm max}$) after MUCA production runs are listed in TABLE~II, 
and the values are larger than $0.8$ in all systems.
We remark that the flatness criteria for our WL simulations was $0.5$.
It means that our estimate of DOS by the REWL-MUCAREM protocol is very accurate indeed.
Similar results were obtained in all system sizes.

The tunneling events during the MUCA production runs were also counted.
Here, a tunneling event is defined by a trajectory that goes from $E_{\rm min}$ to $E_{\rm max}$ and back
(or goes from $E_{\rm max}$ to $E_{\rm min}$ and back).
TABLE~II lists the total number of tunneling events of $32$ independent MUCA production runs.
A lot of tunneling events were indeed observed in all system sizes.
It implies that the observed configurations changed dramatically during the simulation many times.
We concluded that our REWL-MUCAREM protocol and MUCA production run worked properly from these results.

Our estimates of $W_{0}$ are also listed in TABLE~II.
The values obtained from Eq.~(\ref{E_ERROR}) and the extrapolation are shown in Fig.~\ref{RES_ENT}.
We used the following form as an extrapolation formula:
\begin{eqnarray}
   \label{E19}
   \displaystyle
 W_{0} \left(\frac{1}{N} \right) = W_{0}(0) + a \left(\frac{1}{N} \right)^{\theta}.
\end{eqnarray}
Here, we have $\theta \neq 1$ reflects bond correlations in the ground state \cite{ICE_BERG_2007}.
The final estimate of $W_{0}^{{\rm This Work}}$ (which is equal to $W_{0}(0)$ in Fig.~\ref{RES_ENT}) 
is given in the last row of TABLE~III. 
The data points for smaller lattice sizes are included in the fit, 
but not shown in Fig.~\ref{RES_ENT} because we would like to focus on the large lattice $N$ region. 
The final estimate is
\begin{eqnarray}
   \label{E20}
   \displaystyle
 W_{0}(0)=1.507412 \pm 0.000047.
\end{eqnarray}
This estimate converts into 
\begin{eqnarray}
   \label{E21}
   \displaystyle
 S_{0} = 0.815538 \pm 0.000062~{\rm [cal/deg~mole]}.
\end{eqnarray}
The parameters of the fit is also consistent and their values are $a=1.944138 \pm 0.04603$ and $\theta =0.912278 \pm 0.006532$. 

We would like to compare our latest estimate of $W_{0}$ with the results of other research groups.
In Fig.~\ref{COMPARE}, the estimation values of $W_{0}$ with their error bar were plotted.
Various calculation methods  for $S_{0}$ and $W_{0}$ and their calculated values were summarized in TABLE~III.
The relative error between our result and the estimate of Nagle is $0.038$ \%.
We used the following formula as relative error.
\begin{eqnarray}
   \label{E22}
   \displaystyle
 \varepsilon = \frac{\left| A-A_{0} \right|}{A_{0}}.
\end{eqnarray}
Here, $A$ is our measured value and $A_{0}$ is Nagle's theoretical estimate.
Our previous evaluation in $2012$ \cite{ICE_BERG_2012} by MUCA showed that the difference was $0.017$ \%.
However, we considered that our latest estimate is more reliable than that of previous one because of the accuracy of the random number generator.
The Metropolis criteria based on MUCA weight factor in Eq.~(\ref{E12}) might not have worked properly
in large systems (especially, the system for $N=2880$: see Appendix A) in \cite{ICE_BERG_2012}.
Our latest estimate is within the error of the estimates by MUCA simulation in $2007$ \cite{ICE_BERG_2007}, 
in which the problem of random number generator did not occurred.
In order to estimate the residual entropy with higher accuracy than our latest results, 
the calculation of $W_{0}$ on systems larger than $N=4704$ will be necessary.
Although our latest results are slightly different from our previous results in \cite{ICE_BERG_2012}, 
three different computational approaches (PEPS algorithm \cite{ICE_VANDERSTRAETEN}, 
Thermodynamic Integration \cite{ICE_KOLAFA} and REWL-MUCAREM) give almost the same estimates.

\section{conclusions}
Although the theoretical or experimental estimate is still difficult,
the residual entropy of ice Ih is becoming one of good models for testing the accuracy of simulation algorithms 
because of the rapid computational development in recent years. 
However, there seem to be small disagreements among the results of these simulations.
The exact residual entropy of Ice Ih has yet to be obtained.
In this article, we estimated the residual entropy by the REWL-MUCAREM simulations.
Although our final estimate is slightly different from that of the previous MUCA simulation in \cite{ICE_BERG_2012},
it agreed well with the results of several simulation groups and 
three different computational groups gave almost same estimates.
We also discussed the importance of the uniformity of pseudo random number generators in Appendix A.
i

The REWL-MUCAREM strategy can be useful to estimate DOS with high accuracy for the systems 
which have rough energy landscapes, for example, spin-glass or protein systems.
By combining with the reweighting techniques, more information about the systems can be obtained in detail.
In addition, REWL-MUCAREM protocol can also be used in molecular dynamics (MD) simulations.
The problem of discrete random numbers in MC simulations can be avoid by MD simulations.
(Perhaps, {\it Statistical temperature molecular dynamics method} (STMD) \cite{STMD,RESTMD} 
or {\it meta-dynamics algorithm}  \cite{META1,META2,META3}, 
which has a close relationship to WL, is proper to the systems.)
In this case, we can incorporate many techniques which improve the efficiency of sampling 
(e.g., \cite{RESTMD_META}) into REWL-MUCAREM MD.
We hope that the REWL-MUCAREM strategy will give us more reliable insights into complex systems.

\leftline{{\bf Acknowledgements:}}
Some of the computations were performed on the supercomputers at the Supercomputer Center, Institute for Solid State Physics, University of Tokyo.

\newpage
\leftline{{\bf Appendix A: The Effects of Random Numbers on Multicanonical Monte Carlo Simulations}}
\setcounter{figure}{0}
\setcounter{table}{0}
\setcounter{equation}{0}
\setcounter{section}{1}
\renewcommand{\thefigure}{\Alph{section}\arabic{figure}}
\renewcommand{\thetable}{\Alph{section}-\Roman{table}}
\renewcommand{\theequation}{\Alph{section}\arabic{equation}}

There is no doubt that the quality of pseudo random number generators strongly affects 
the results of Monte Carlo simulations. 
Pseudo random number generators have their own characteristics, for example, periodicity of random numbers.
Here, we would like to discuss the minimum value which can be generated by random number generators and the effects on the MUCA MC simulations.

We used two well-known pseudo random number generators, namely, 
Marsaglia pseudo random number generator \cite{MARSAGLIA1} and Mersenne Twister pseudo random number generator \cite{MERSENNE1}.
Marsaglia generator was employed in our previous studies \cite{ICE_BERG_2007,ICE_BERG_2008,ICE_BERG_2012}.
Mersenne Twister generator was used in this work.
The source codes are found in \cite{MUCA_BOOK,MERSENNE_CODE}. 

In order to compare the accuracy of random numbers, 
pseudo random numbers were generated $10^{11}$ times by these generators.
The generated values less than $5.0 \times 10^{-7}$ by Marsaglia generator (green dots) 
and Mersenne Twister generator (purple dots) are plotted in Fig.~\ref{RANDOM_NUMBERS}.
Although random numbers by Mersenne Twister generator seems to make a uniform distribution,
we can see a discrete distribution by Marsaglia generators. 
Thus, samples by Marsaglia make green lines in Fig.~\ref{RANDOM_NUMBERS}.
The minimum random number value by Marsaglia generator was $0$ 
and the next minimum value was $5.9605 \times 10^{-8}$.
The random seeds were Seed1$=11$ and Seed$2=20$. 
It means that Marsaglia generator we employed cannot generate the values within $(0,5.9605 \times 10^{-8})$ as a random number.
On the other hand, the minimum random number value by Mersenne Twister generator (the random seed is $4357$) was $0$
and the next minimum value was $2.3283 \times 10^{-10}$ in our test, which is smaller 
than the value $5.9605 \times 10^{-8}$ by Marsaglia.

In the 2-state model, the transition probability $w(X_{0} \rightarrow X_{1})=\exp(-\Delta S)$, where $\Delta S=\ln n_{1}-\ln n_{0}$,
during MUCA simulations from the ground state $X_{0}$
to the first excited state $X_{1}$ are shown in Fig.~\ref{TRANSITION}.
The inset in Fig.~\ref{TRANSITION} shows the differences of the estimate of entropy $\Delta S$ 
between the ground state (the value of entropy is $\ln n_{0}$) and the first exited state (the value of entropy is $\ln n_{1}$).
It is clear that the difference becomes larger as the number of molecules increases.
Thus, the acceptance probability around the ground state becomes small.
The $w(X_{0} \rightarrow X_{1} )$ is approximately to $e^{-16.0} (\simeq 1.125 \times 10^{-7})$ for $N = 4704$.
The Marsaglia generator would not work properly because of the badness of the uniformity of random numbers.
In addition, we might not have obtained a proper estimate for $N=2880$ in our previous work in \cite{ICE_BERG_2012}. 
This is the reason why our latest estimate of residual entropy ($S_{0} = 0.815538 \pm 0.000062~{\rm [cal/deg~mole]}$) 
in this article is different from our previous result ($S_{0} = 0.815148 \pm 0.000047~{\rm [cal/deg~mole]}$).
Note that there are a sophisticated Marsaglia random number generator to alleviate the discrete problem
by combining two Marsaglia random numbers into one \cite{MUCA_BOOK}.

\newpage
\setcounter{figure}{0}
\setcounter{table}{0}
\setcounter{equation}{0}
\setcounter{section}{1}
\renewcommand{\thefigure}{\arabic{figure}}
\renewcommand{\thetable}{\Roman{table}}
\renewcommand{\theequation}{\arabic{equation}}
\begin{table}[H]
\begin{threeparttable}[h]
\caption{\label{tab:table1} Initial conditions in REWL-MUCAREM simulations.
}
\begin{tabular*}{16.5cm}{@{\extracolsep{\fill}}ccccccccc}
 \hline 
 \hline 
 $N$ &
 ~~~$n_{x}$  &
 $n_{y}$  &
 $n_{z}$  &
 ~~~No. of replicas &
 ~~~Replica Exchange\tnote{a} &
 ~~~WL criteria\tnote{b} &
 ~~~Total MC sweeps &
 ~~~Total MC sweeps  \rule[0mm]{0mm}{4mm}\\
  &
  &
  &
  &
  &
  &
  &
 for REWL \tnote{c} &
 ~~~for MUCAREM \tnote{d} \\ 
 \hline 
 $128$ &
 $4$ &
 $8$ &
 $4$ &
 $4$ &
 ~~$250$ &
 ~~$500$ &
 $1.350 \times 10^{5}$ &
 ~$2.0 \times 10^{7}  \times 5$ \rule[0mm]{0mm}{4mm}\\

 $288$ &
 $4$ &
 $12$ &
 $6$ &
 $8$ &
 ~~$250$ &
 ~~$500$ &
 $4.280 \times 10^{5}$ &
 $2.0 \times 10^{7}  \times 5$ \\

 $360$ &
 $5$ &
 $12$ &
 $6$ &
 $8$ &
 ~~$250$ &
 ~~$500$ &
 $4.185 \times 10^{5}$ &
 $2.0 \times 10^{7} \times 5$  \\

 $576$ &
 $6$ &
 $12$ &
 $8$ &
 $16$ &
 ~~$500$ &
 ~~$1000$ &
 $6.310 \times 10^{5}$ &
 $2.0 \times 10^{7} \times 5$  \\

 $896$ &
 $7$ &
 $16$ &
 $8$ &
 $16$ &
 ~~$500$ &
 ~~$1000$ &
 $1.408 \times 10^{6}$ &
 $2.0 \times 10^{7} \times 5$ \\

 $1600$ &
 $8$ &
 $20$ &
 $10$ &
 $32$ &
 ~~$2500$ &
 ~~$5000$ &
 $2.095 \times 10^{6}$ &
 $2.0 \times 10^{7}  \times 5$ \\

 $2880$ &
 $10$ &
 $24$ &
 $12$ &
 $32$ &
 ~~$2500$ &
 ~~$5000$ &
 $7.065 \times 10^{6}$ &
 $2.0 \times 10^{7}  \times 5$ \\

 $4704$ &
 $12$ &
 $28$ &
 $14$ &
 $32$ &
 ~~$5000$ &
 ~~$10000$ &
 $1.405 \times 10^{7}$ &
 $2.0 \times 10^{7} \times 5$  \\

 \hline 
 \hline 
\end{tabular*}

\begin{tablenotes}
\item[a] The interval of replica exchange trial (MC sweeps) in REWL and MUCAREM.
\item[b] The interval of WL criteria check (MC sweeps) in REWL.
\item[c] Total MC sweeps per each replica that is required for all WL weight factors $f$ to converge to $f_{\rm final}$ in REWL.
\item[d] Total MC sweeps per each replica in MUCAREM. MUCAREM simulations were repeated $5$ times.
\end{tablenotes}

\end{threeparttable}
\end{table}
\begin{table}[H]
\begin{threeparttable}[h]
\caption{\label{tab:table2} Estimated residual entropy of Ice Ih.
}
\begin{tabular*}{16.5cm}{@{\extracolsep{\fill}}cccccccc}
 \hline 
 \hline 
 ~~~~$N$~~~ &
 $n_{x}$  &
 $n_{y}$  &
 $n_{z}$  &
 ~~Tunneling\tnote{a} &
 ~~Flatness\tnote{b} &
 ~~$W_{0}$\tnote{*} &
 $S_{0} $\tnote{*} \rule[0mm]{0mm}{4mm}\\
 \hline 
 $128$ &
 $4$ &
 $8$ &
 $4$ &
 ~~$7612228$ &
 ~~$0.98391$ &
 ~~$1.5286054(462) $ &
 ~~~~$0.8432816(601) $ \rule[0mm]{0mm}{4mm}\\

 $288$ &
 $4$ &
 $12$ &
 $6$ &
 ~~$1598145$ &
 ~~$0.97176$ &
 ~~$1.5176118(362) $ &
 ~~~$0.8289382(474) $ \\

 $360$ &
 $5$ &
 $12$ &
 $6$ &
 ~~$1020866$ &
 ~~$0.97870$ &
 ~~$1.5156001(402) $ &
 ~~~$0.8263023(527) $ \\

 $576$ &
 $6$ &
 $12$ &
 $8$ &
 ~~$404617$ &
 ~~$0.97047$ &
 ~~$1.5127892(339) $ &
 ~~~$0.8226133(446) $\\

 $896$ &
 $7$ &
 $16$ &
 $8$ &
 ~~$172052$ &
 ~~$0.95956$ &
 ~~$ 1.5109753(276) $ &
 ~~~$ 0.8202291(363) $\\

 $1600$ &
 $8$ &
 $20$ &
 $10$ &
 ~~$ 57171$ &
 ~~$ 0.93313$ &
 ~~$ 1.5095170(284) $ &
 ~~~$ 0.8183102(373) $\\

 $2880$ &
 $10$ &
 $24$ &
 $12$ &
 ~~$ 17417$ &
 ~~$ 0.90334$ &
 ~~$ 1.5086586(304) $ &
 ~~~$ 0.8171799(401) $ \\

 $4704$ &
 $12$ &
 $28$ &
 $14$ &
 ~~$ 6318$ &
 ~~$ 0.83998$ &
 ~~$ 1.5082141(319) $ &
 ~~~$ 0.8165944(420)$ \\

 $ \infty $&
 $ $ &
 fitting &
 $ $ &
   &
   &
 ~~$ 1.5074123(466) $ &
 ~~~$ 0.8155376(614) $ \\
 \hline 
 \hline 
\end{tabular*}

\begin{tablenotes}
\item[a] The total counts of observed tunneling events during 32 MUCA production runs.
\item[b] The value of flatness ($H_{\rm max} / H_{\rm min}$) after 32 MUCA production runs.
\item[*] The values in parentheses represent the errors obtained by $32$ MUCA production runs and fitting, using Eq. (\ref{E_ERROR}).
\end{tablenotes}

\end{threeparttable}
\end{table}
\begin{table}[H]
\caption{\label{tab:table3} Comparing the estimates of various methods.
}
\begin{tabular*}{16.5cm}{@{\extracolsep{\fill}}llllll}
 \hline 
 \hline 
 Group    &
 Methods &
 $W_{0}$ &
 $\Delta W_{0}$ &
 $S_{0}$ &
 $\Delta S_{0}$  \rule[0mm]{0mm}{4mm} \\
 \hline 
 Nagle \cite{ICE_NAGLE}&
 Series expansion &
 $1.50685 $ &
 $0.00015$ &
 $0.8147962 $ &
 $0.000198$
   \rule[0mm]{0mm}{4mm} \\
 Berg (2007) \cite{ICE_BERG_2007}&
 Multicanonical algorithm&
 $1.50738 $ &
 $0.00016$ &
 $0.81550 $ &
 $0.00021$
  \\
 Berg (2012) \cite{ICE_BERG_2012}&
 Multicanonical algorithm&
 $1.507117 $ &
 $0.000035$ &
 $0.815149 $ &
 $0.000046$
  \\
 Herrero \cite{ICE_HERRERO}&
 Thermodynamic Integration &
 $1.50786 $ &
 $0.00012$ &
 $0.81613 $ &
 $0.00016$
  \\
 Kolafa \cite{ICE_KOLAFA}&
 Thermodynamic Integration &
 $1.5074674 $ &
 $0.0000038$ &
 $0.8156103 $ &
 $0.0000051$
  \\
 Ferreyra \cite{ICE_FERREYRA2}&
 Wang-Landau algorithm &
 $1.5070 $ &
 $0.0009$ &
 $0.81478 $ &
 $0.00012$
  \\
 Vanderstraeten \cite{ICE_VANDERSTRAETEN}&
 PEPS algorithm&
 $1.507456$ &
 $$ &
 $0.8155953$
 $$
  \\
 \hline 
 This work &
 REWL-MUCAREM &
 $1.507412$ &
 $0.000047$ &
 $0.815538$ &
 $0.000062$
   \\
 \hline 
 \hline 
\end{tabular*}

\end{table}

\begin{figure}[H]
\begin{tabular}{ccc}
\begin{minipage}[t]{0.45\textwidth}
  \centering
 \includegraphics[width=7.5cm,height=6.0cm]{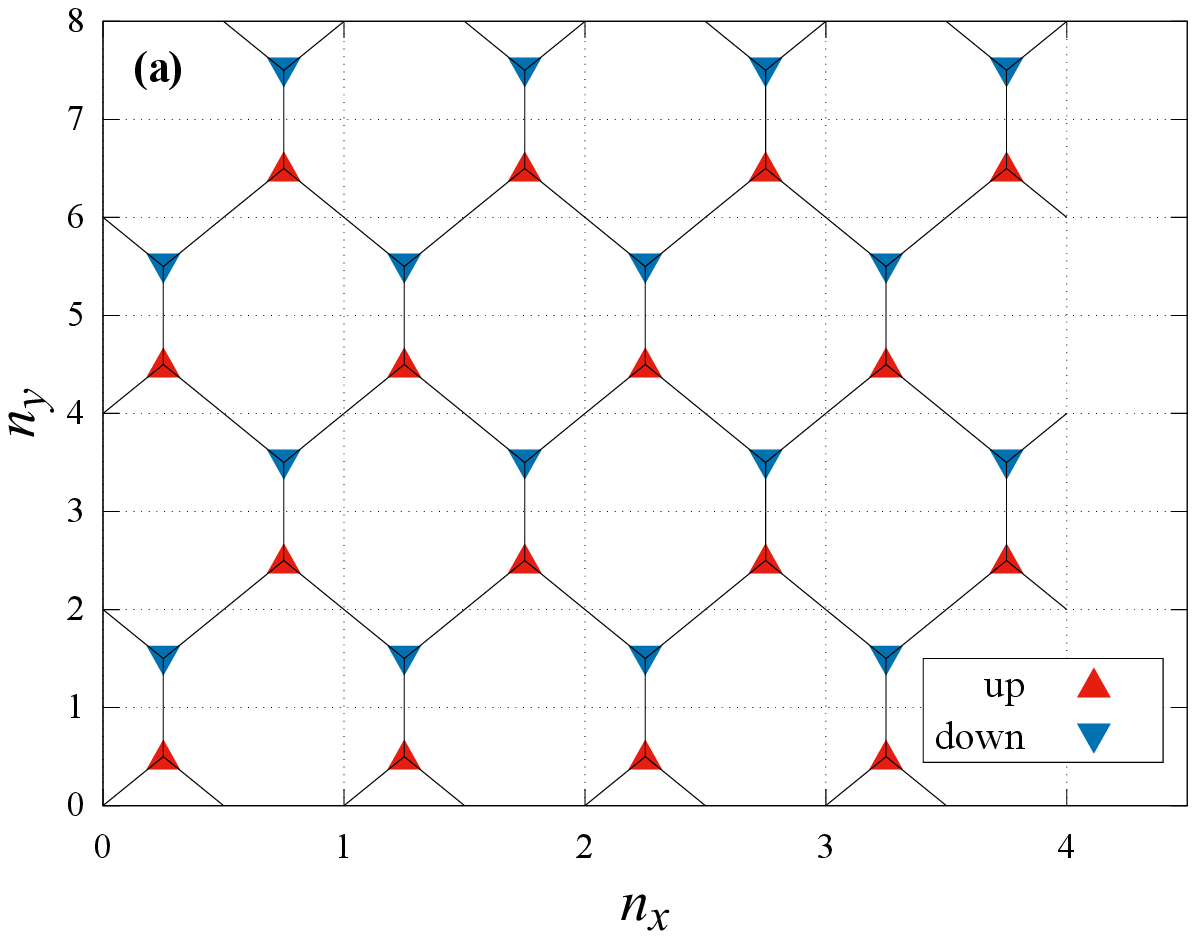}
\end{minipage}

\begin{minipage}[t]{0.05\textwidth}
\hspace{0.1cm}
\end{minipage}

\begin{minipage}[t]{0.45\textwidth}
  \centering
 \includegraphics[width=7.5cm,height=6.0cm]{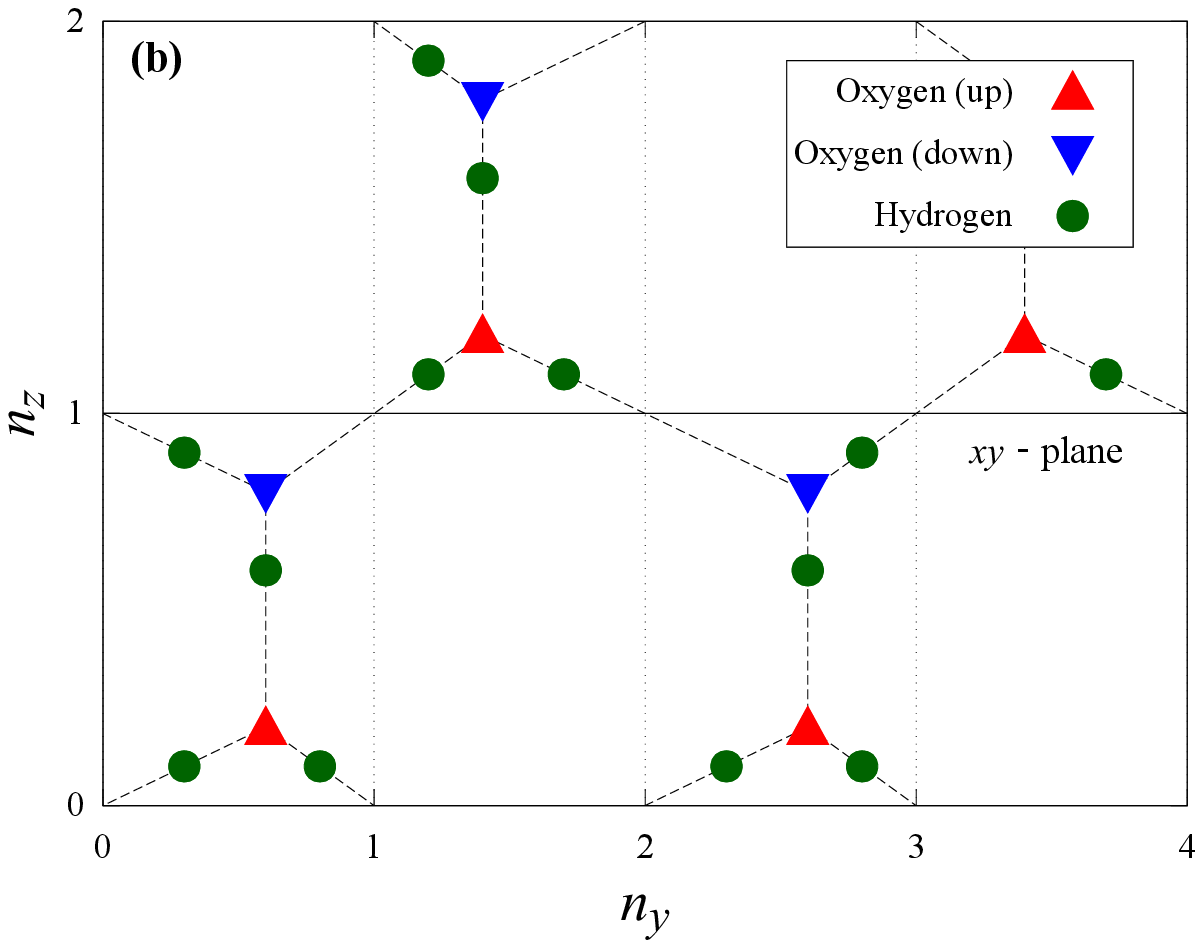}
\end{minipage}
\end{tabular}

  \caption{Two-dimensional projection of ice Ih. 
(a) shows the projection to the $xy$-plane and (b) shows the projection to the $yz$-plane.
The scale is different from the actual ice Ih structure for simplicity. 
$n_{x},n_{y}$, and $n_{z}$ are the numbers of sites along the $x,y,$ and $z$ axis, respectively.
The total number of water molecules $N$ is given by $n_{x} \times n_{y} \times n_{z}$.
The red triangles imply that the lattice points exist above the $xy$-plane, and
the blue triangles imply that the lattice points exist below the $xy$-plane in (a).
Oxygen atoms are located on lattice points.
The triangles in (b) also represent the oxygen atoms. 
The dotted lines represent the hydrogen bonds pair of oxygen atoms.
The filled green circles are hydrogen atoms on chemical bonds.
Hydrogen atoms can occupy one of the two places on each bond according to the ice rules.}
  \label{STR_2D}
\end{figure}
\begin{figure}[H]
\begin{tabular}{ccc}
\begin{minipage}[t]{0.45\textwidth}
  \centering
  \includegraphics[width=7.5cm,height=6.0cm]{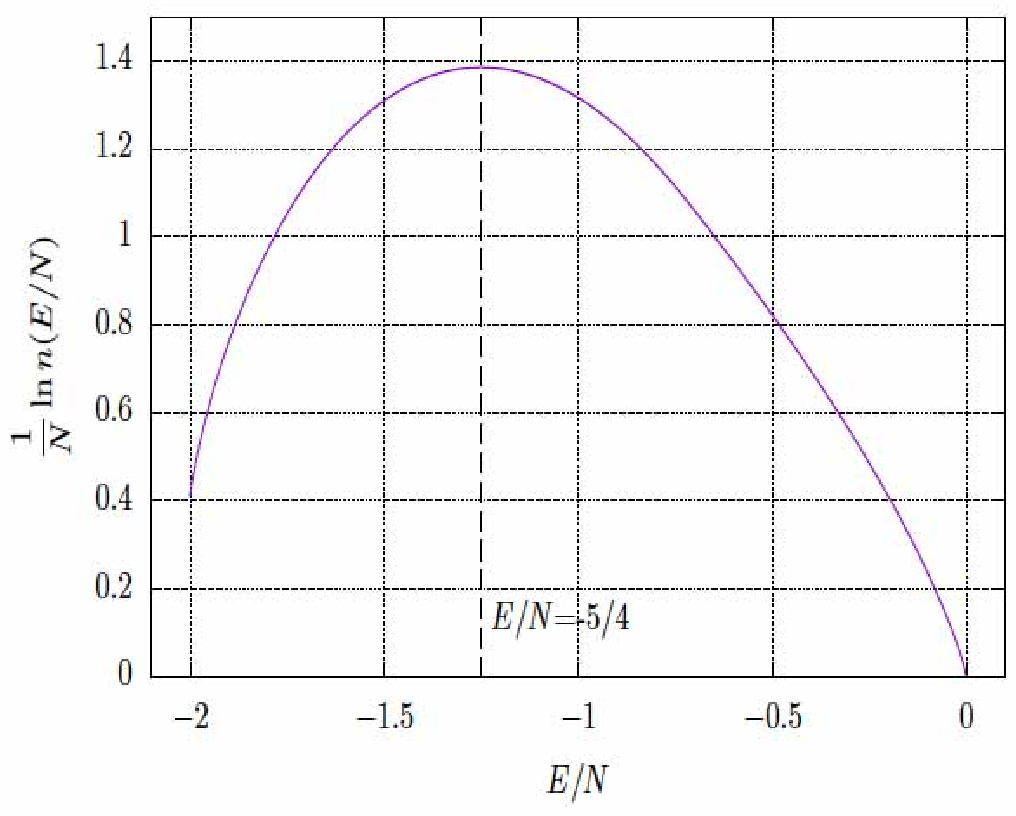} 
  \caption{Typical dimensionless entropy $\ln n(E/N)$ per one water molecule of 
  ice Ih as a function of potential energy per site $(E/N)$.
 The values were obtained by additional REWL-MUCAREM simulation for $N=2880$ system.
 $\ln n(0)$ is set to $\ln (2)$ because the possible configures at $E=0$ are $2$.
 The entropy takes the maximum value at the energy $E/N=-5/4$.}
  \label{TPC_ENT}
\end{minipage}

\begin{minipage}[t]{0.05\textwidth}
\hspace{0.1cm}
\end{minipage}

\begin{minipage}[t]{0.45\textwidth}
  \includegraphics[width=7.5cm,height=6.0cm]{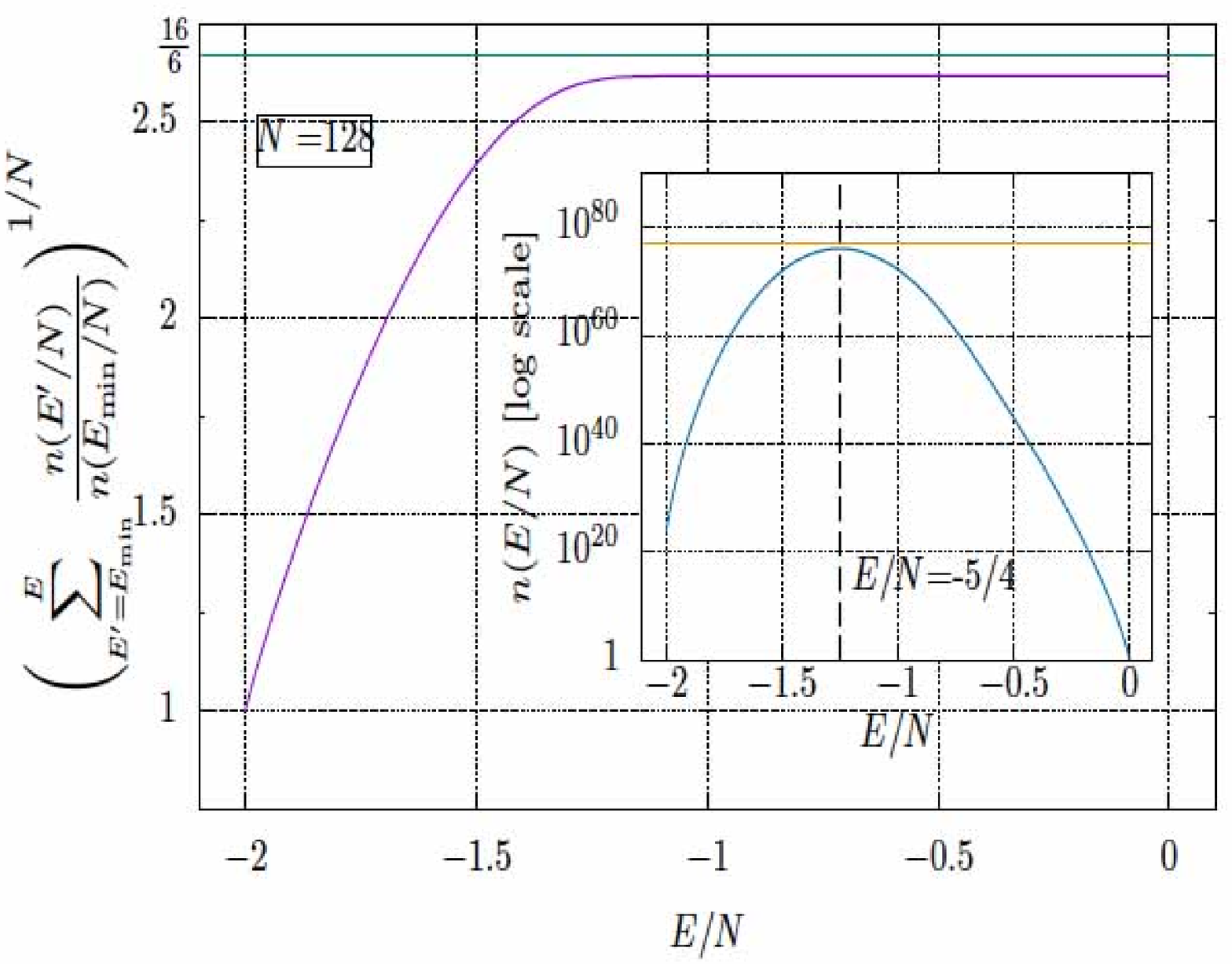}
\caption{The summation of $n(E/N)$ from $E_{\rm min}$ to $E$ for the system $N=128$.
 The value is normalized at $E_{\rm min}$ per one water molecule.
 The horizontal green line shows the inverse of Pauling's estimate $(6/16)$.
 The summation is saturated around a bit larger potential energy than $E/N=-5/4$.
 The inset shows the $n(E/N)$ we obtained. Here, $n(0)$ is set to $2$.
 The horizontal orange line shows the total number of configurations $(\sum_{E} n(E/N) = 2^{2N})$.
 $n(E/N)$ takes the maximum value at $E=-5/4$ and 
 most of the total number of conformations are distributed around the peak.} 
 \label{TPC_NOS}
\end{minipage}
\end{tabular}
\end{figure}

\begin{figure}[H]
\begin{tabular}{ccc}
\begin{minipage}[t]{0.45\textwidth}
  \centering
 \includegraphics[width=7.5cm,height=6.0cm]{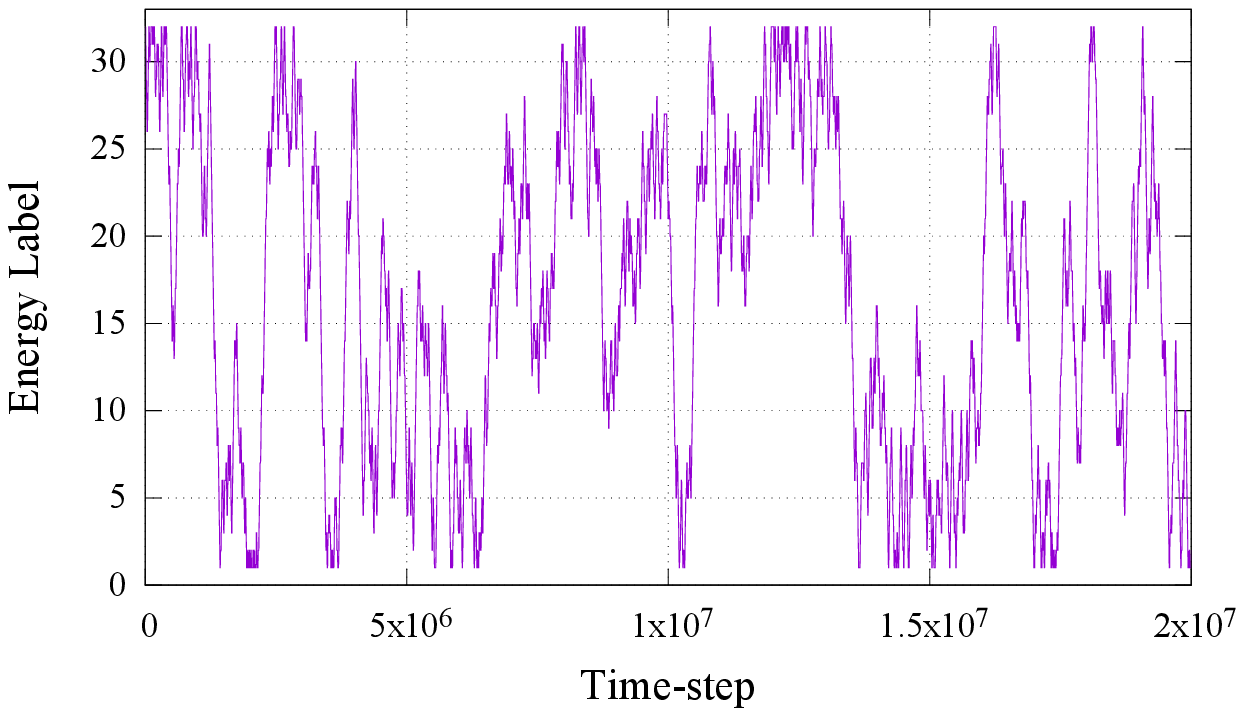}
  \caption{History of the energy-range index (Energy label) of one of the replicas (Replica 1) during the final MUCAREM simulation 
for $N=4704$.} 
  \label{MCRM_ERANK}
\end{minipage}

\begin{minipage}[t]{0.05\textwidth}
\hspace{0.1cm}
\end{minipage}

\begin{minipage}[t]{0.45\textwidth}
 \includegraphics[width=7.5cm,height=6.0cm]{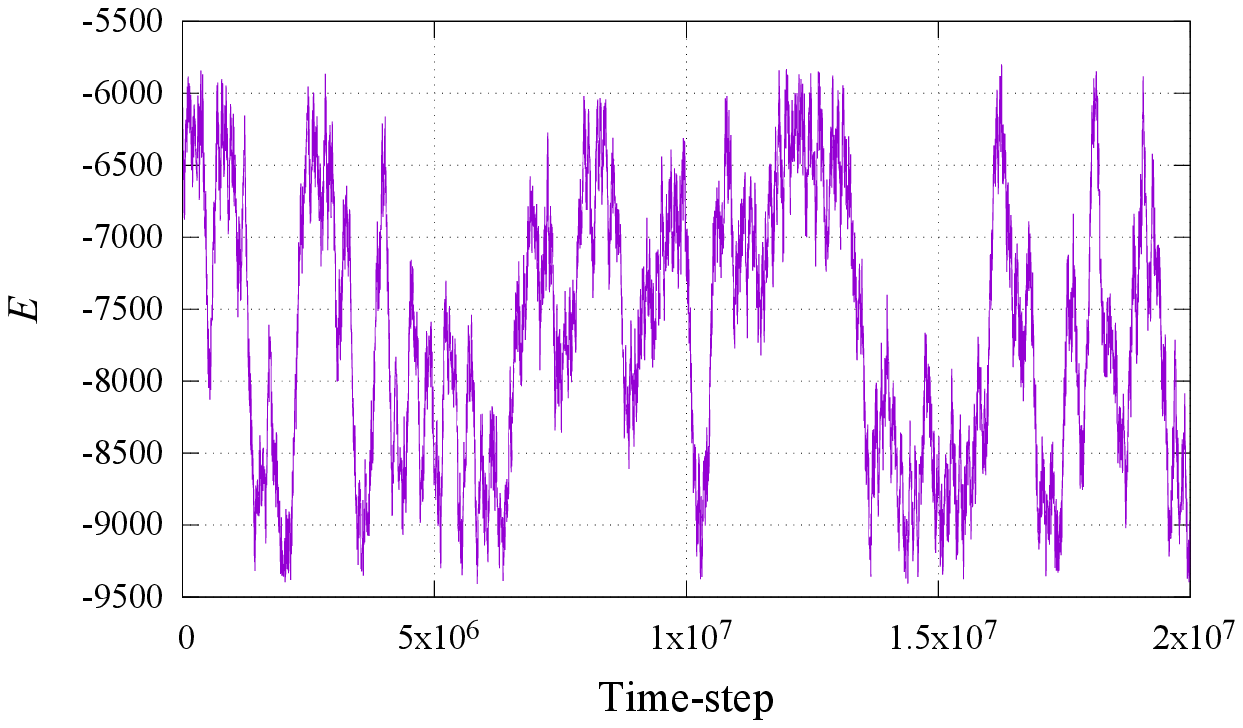}
  \caption{History of the potential energy of one of the replicas (Replica 1) during the final MUCAREM simulation for  $N=4704$. } 
  \label{MCRM_E}
\end{minipage}
\end{tabular}
\end{figure}
\begin{figure}[H]
  \centering
 \includegraphics[width=12.0cm]{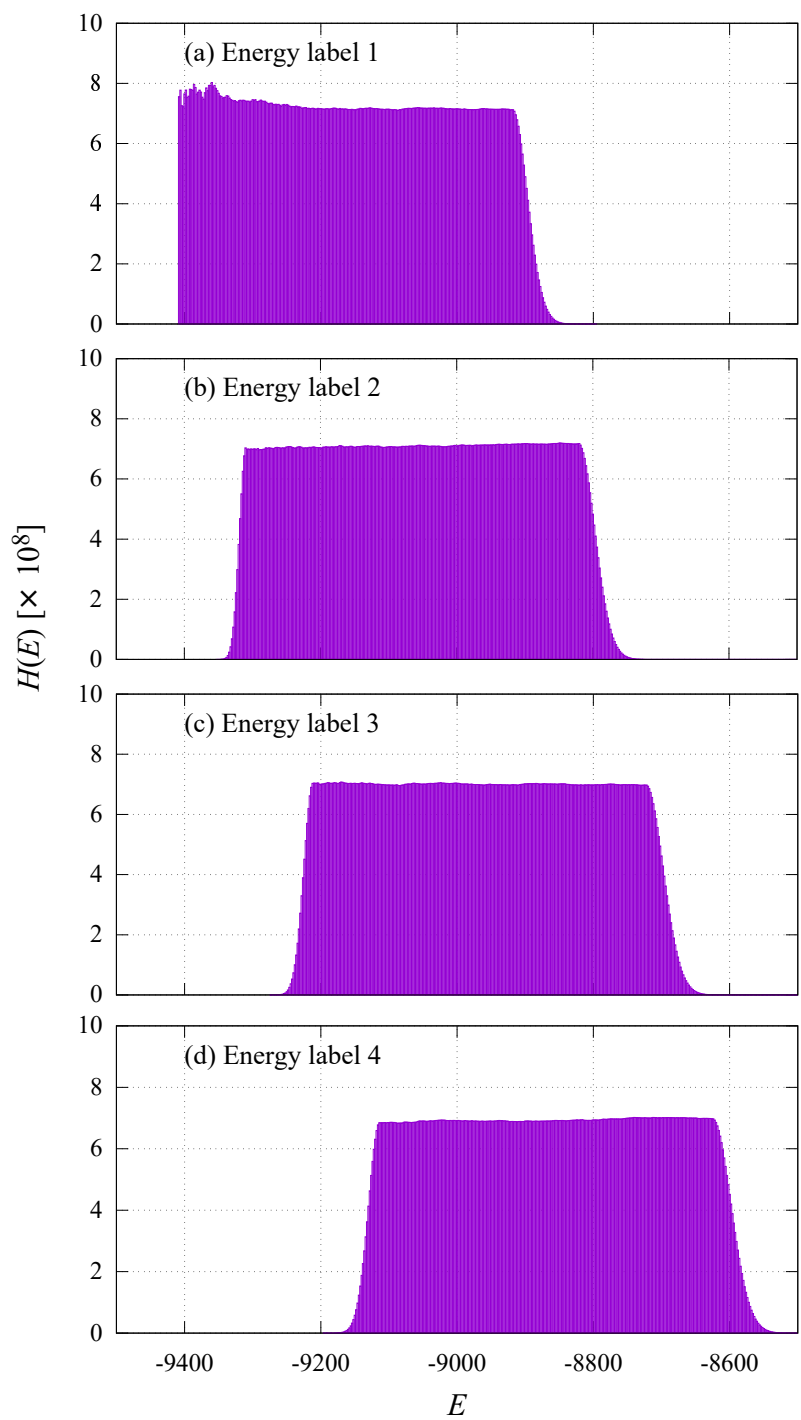}
  \caption{Histograms of potential energy obtained by the final MUCAREM simulation of the water molecules  $N=4704$.
 Each energy label corresponds to the sub-region $m$. 
 Sub-regions have an overlap of about $80$ \% between neighboring sub-regions.
 Each histogram shows a flat distribution.}
  \label{MCRM_HIST}
\end{figure}
\begin{figure}[H]
\begin{tabular}{ccc}
\begin{minipage}[t]{0.45\textwidth}
  \centering
 \includegraphics[width=7.5cm,height=6.0cm]{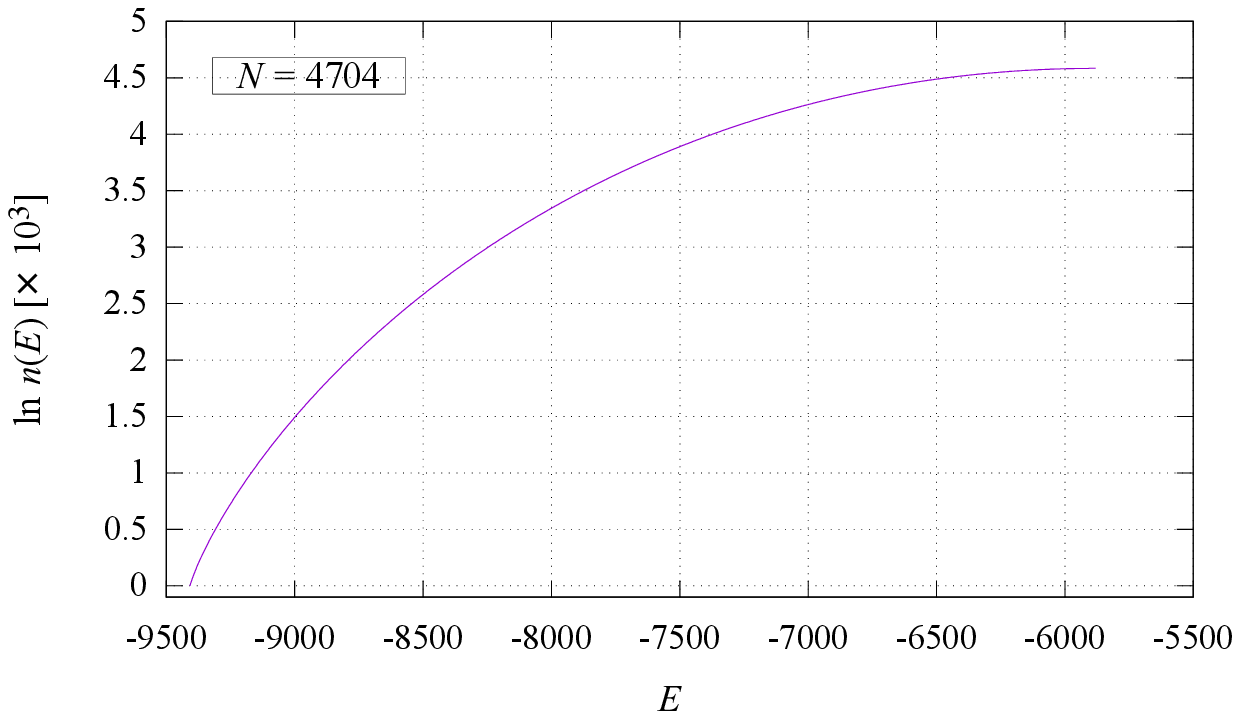}
  \caption{The entropy as a function of energy $E$ estimated by the REWL-MUCAREM simulation for $N=4704$.
  Here, the value of $\ln n(E)$ at $E=-9408$ is set equal to $0$.} 
  \label{MUCA_ENT}
\end{minipage}

\begin{minipage}[t]{0.05\textwidth}
\hspace{0.1cm}
\end{minipage}

\begin{minipage}[t]{0.45\textwidth}
 \includegraphics[width=7.5cm,height=6.0cm]{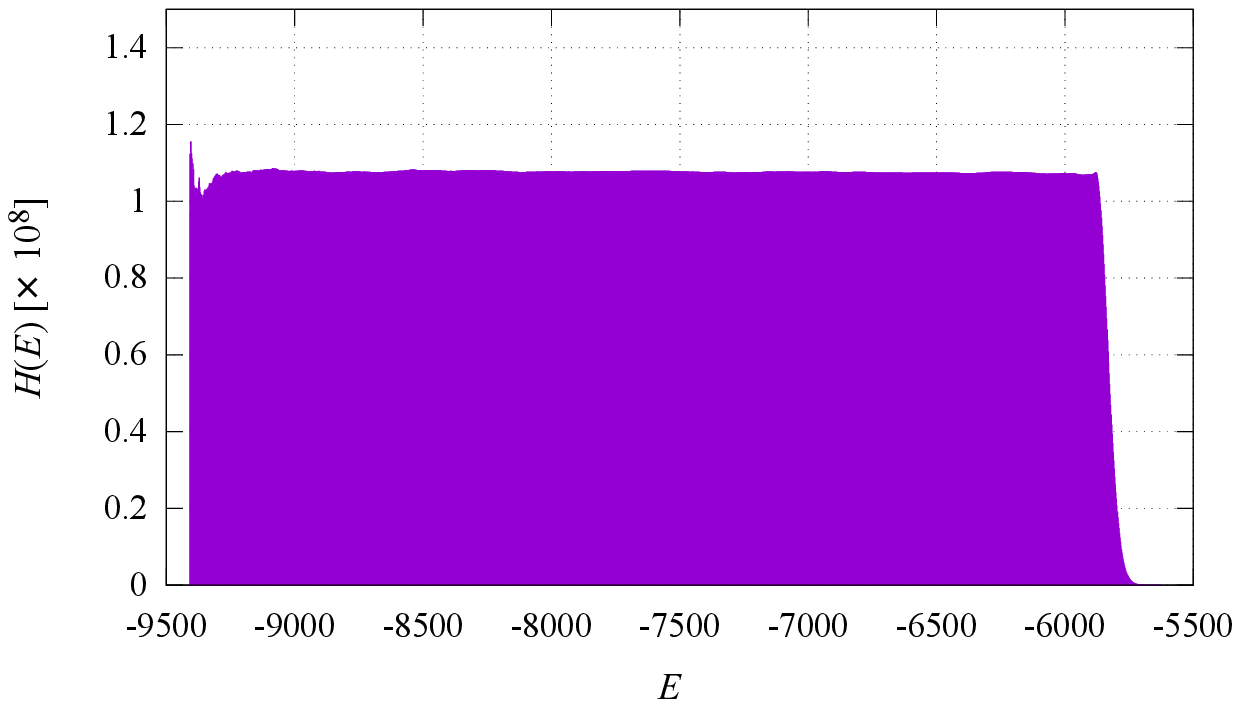}
  \caption{Total histogram of potential energy obtained by the MUCA production simulation for $N=4704$.
  The entropy in Fig.~\ref{MUCA_ENT} was used as the MUCA weight factor.} 
  \label{MUCA_HIST}
\end{minipage}
\end{tabular}
\end{figure}

\begin{figure}[H]
  \centering
 \includegraphics[width=12.0cm]{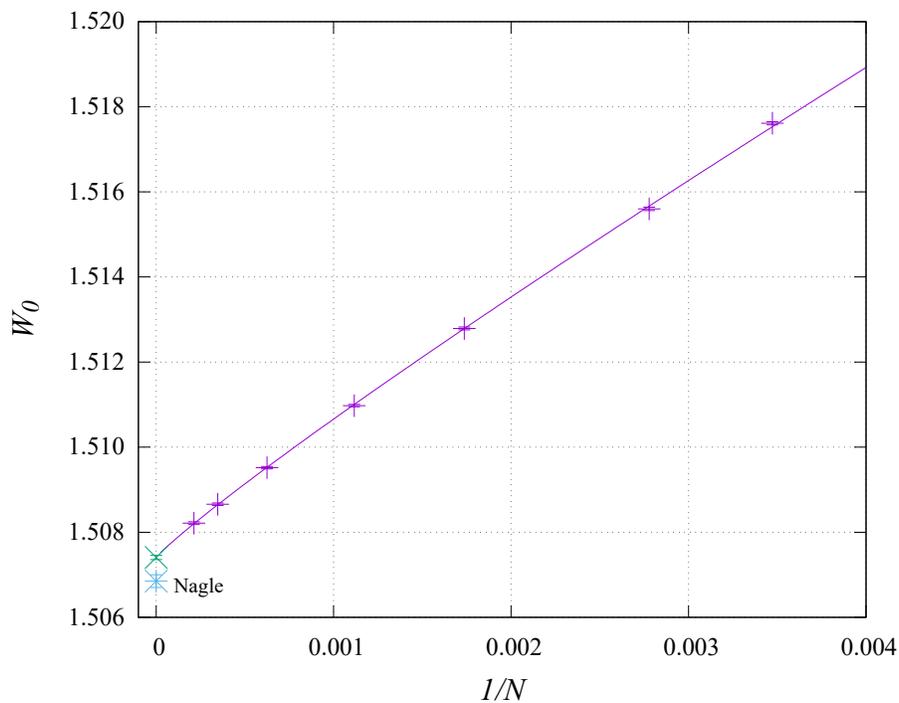}
  \caption{The degree of freedom of the orientation of one water molecule $W_{0}(1/N)$ at ground state as the function of the inverse of $N$. Error bars are smaller than the symbols.} 
  \label{RES_ENT}
\end{figure}
\begin{figure}[H]
  \centering
 \includegraphics[width=12.0cm]{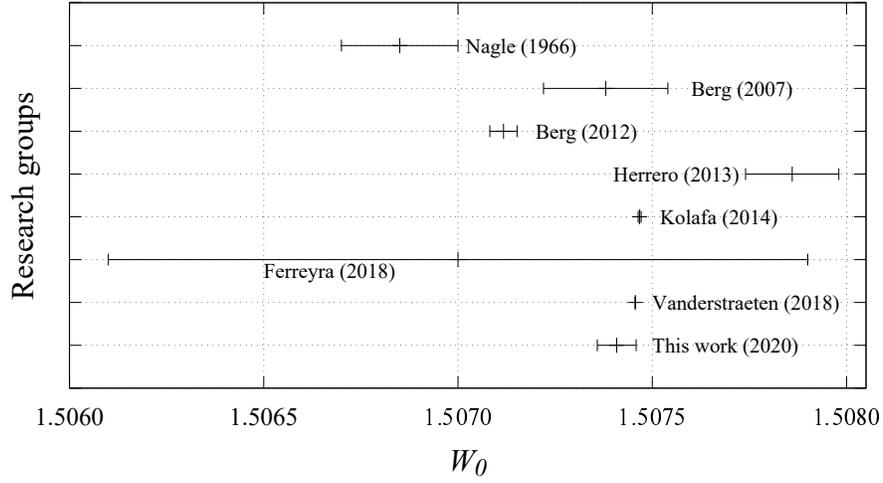}
  \caption{Evaluates of $W_{0}$ by several research groups.} 
  \label{COMPARE}
\end{figure}

%
\setcounter{figure}{0}
\setcounter{table}{0}
\setcounter{equation}{0}
\setcounter{section}{1}
\renewcommand{\thefigure}{\Alph{section}\arabic{figure}}
\renewcommand{\thetable}{\Alph{section}-\Roman{table}}
\renewcommand{\theequation}{\Alph{section}\arabic{equation}}
\begin{figure}[H]
\begin{tabular}{ccc}
\begin{minipage}[t]{0.45\textwidth}
\centering
\includegraphics[width=7.5cm,height=6.0cm]{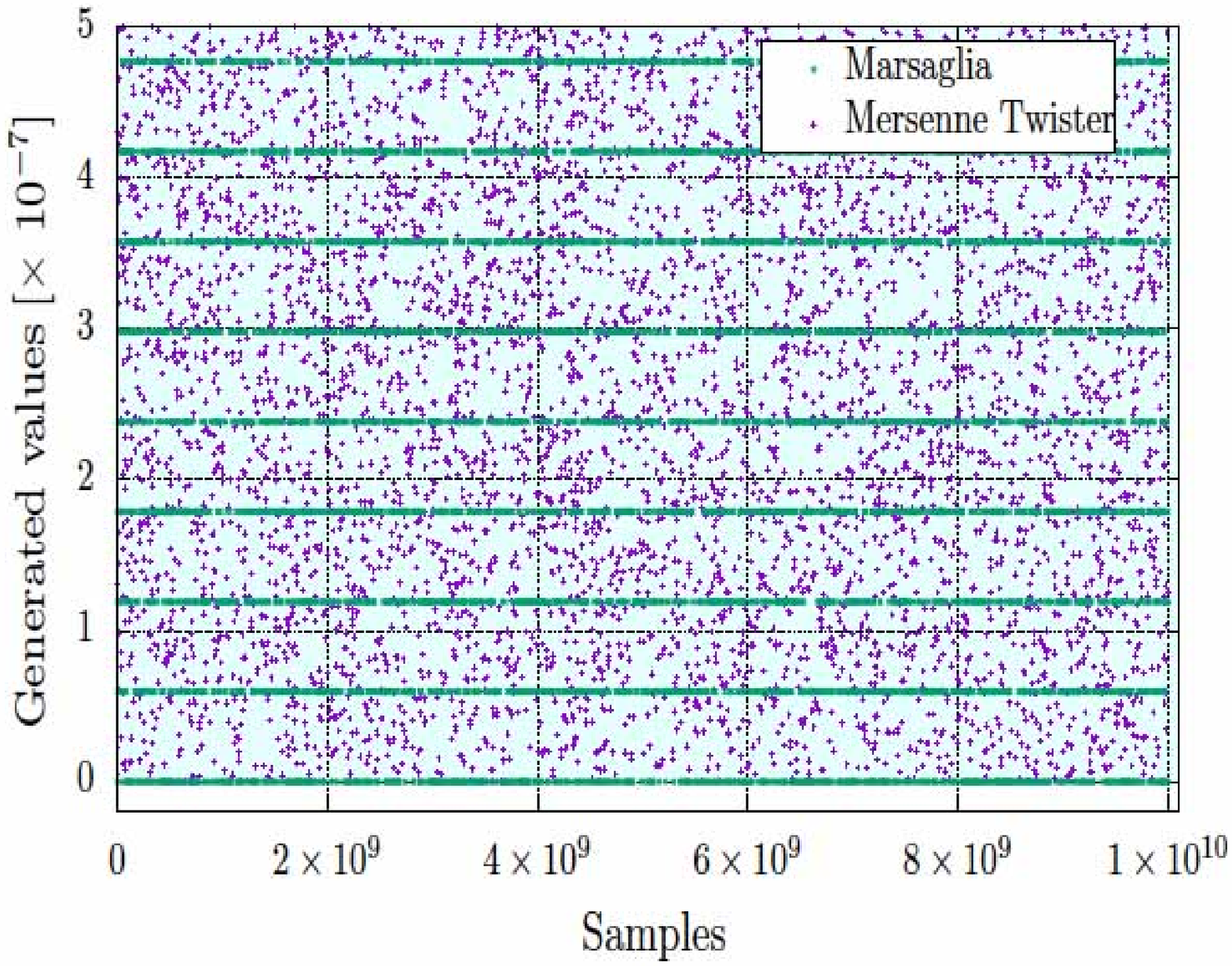}
\caption{Generated random numbers by Marsaglia generator (green) and Mersenne Twister generator (purple).
Although the purple dots seem to be distributed uniformly, green dots only take nine discrete values 
($0,0.59605, 1.19210, 1.78815, 2.38420, 2.98025, 3.57630,$ 
$4.17235, 4.76840 $ [$\times 10^{-7}$ ]).}
\label{RANDOM_NUMBERS}
\end{minipage}

\begin{minipage}[t]{0.05\textwidth}
\hspace{0.1cm}
\end{minipage}

\begin{minipage}[t]{0.45\textwidth}
\centering
\includegraphics[width=7.5cm,height=6.0cm]{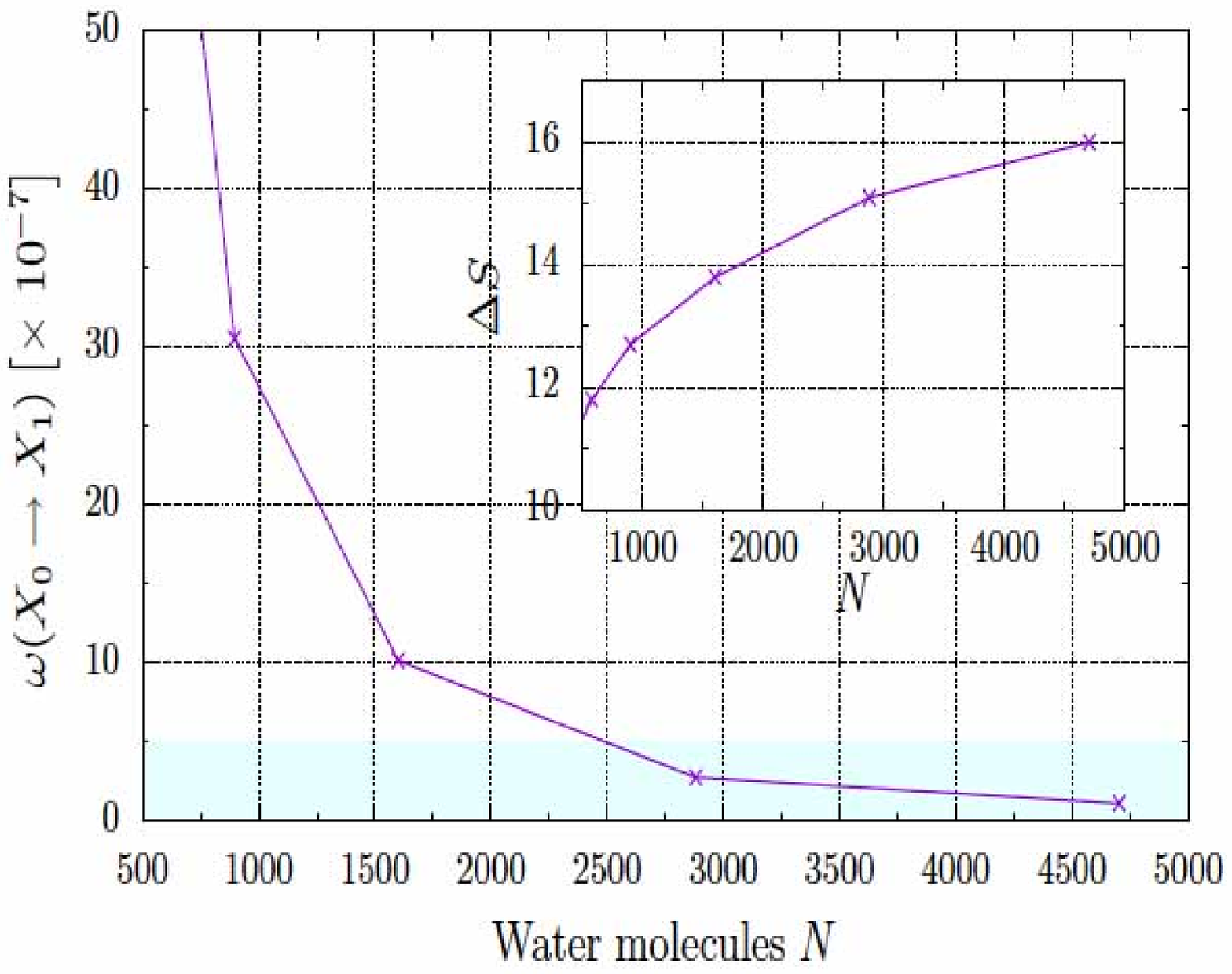}
\caption{Transition probability ($\omega(X_{0} \rightarrow X_{1}) = \exp [- \Delta S]$) from the ground state 
$X_{0}$ (the estimated dimensionless entropy is $\ln n_{0}$) 
to the first excited states $X_{1}$ (the estimated dimensionless entropy is $\ln n_{1}$) in the 2-state model.
Here, $\Delta S$ is defined by $\Delta S = \ln n_{1} - \ln n_{0}$.
The inset shows $\Delta S$. 
The shaded area (light blue region) corresponds to the range of the ordinate in Fig.~\ref{RANDOM_NUMBERS}.}
\label{TRANSITION}
\end{minipage}
\end{tabular}
\end{figure}

\newpage

%
%
\end{document}